\documentclass[aps,pra,twocolumn,nopacs,floatfix]{revtex4}
\usepackage{ulem}
\usepackage{epsfig}
\usepackage{graphicx}
\usepackage{dcolumn}
\usepackage{amsthm,amsmath}
\usepackage{comment}
\usepackage[colorlinks]{hyperref}
\usepackage{xcolor}
\usepackage{mathrsfs}

\usepackage{amsmath}

\usepackage{morefloats}

\begin{document}

\title{Constructing Electron-Atom Elastic Scattering Potentials using Relativistic Coupled-Cluster Theory: A few case studies}

\author{B. K. Sahoo}
\email{bijaya@prl.res.in}

\affiliation{Atomic, Molecular and Optical Physics Division, Physical Research Laboratory, Navrangpura, Ahmedabad 380009, India}

\date{\today}

\begin{abstract}
In view of immense interest to understand impact of an electron on atoms in the low-energy scattering phenomena observed in laboratories and astrophysical processes, we prescribe here an approach to construct potentials using relativistic coupled-cluster (RCC) theory for the determination of electron-atom (e-A) elastic scattering cross-sections (eSCs). The net potential of an electron, scattered elastically by an atom, is conveniently expressed as sum of static ($V_{st}$) and exchange ($V_{ex}$) potentials due to interactions of the scattered electron with the electrons of the atom and potentials due to polarization effects ($V_{pol}$) on the scattered electron by the atomic electrons. The $V_{st}$ and $V_{ex}$ potentials for the e-A eSC problems can be constructed with the knowledge of electron density function of the atom, while the $V_{pol}$ potential can be obtained using polarizabilities of the atom. In this work, we present electron densities and electric polarizabilties 
of Be, Mg, Ne and Ar atoms using two variants of the RCC method. Using these quantities, we construct potentials for the e-A eSC problems. For obtaining $V_{pol}$ accurately, we have evaluated the second- and third-order electric dipole and quadrupole polarizabilities in the linear response 
approach.
\end{abstract}

\maketitle

\section{Introduction}

Accurate estimation of scattering cross-sections of electrons with atomic systems are interest to a wide range of applications in laboratory 
scattering processes and astrophysics \cite{johnson,amusia,Kotlarchyk,anil}. The challenges in the calculations of scattering cross-sections lie in determining
accurate wave functions for the scattered electron in the vicinity of an atomic target \cite{RDressler,Dipti}. The coupling between the 
scattered wave functions and atomic wave functions are taken care through the close-coupling \cite{CC} and R-matrix \cite{Burke} formalism, but
they are mostly used in the non-relativistic framework \cite{Post,jpcdata} owing to the complexity involved in the relativistic formalism. In 
another approach, interactions among the scattered electron and atomic electrons are included by splitting them into two parts -- the electron-electron correlation part and the electron polarization effects due to the atomic electrons \cite{connel,yuan,Franz,tenfen,Felipe,khandker}. In this approach, wave functions of the electron and atom are solved separately. The electron correlation effects within the atom are accommodated via a suitable many-body method in the determination of atomic wave functions (equivalently to atomic wave density functions ($\rho$)). These functions are further used to construct interaction potential for the scattered electron. It has both direct and exchange terms owing to indistinguishably nature of the electrons. Again, an atom is polarized due to the charged scattered electron which modifies the behavior its wave functions. This effect also influences construction of the effective potentials of the scattered electrons and are estimated using electric polarizabilities of the atom. These effective potentials are used to obtain wave functions of the scattered electrons, at different range of kinetic energies, using distorted wave function (DW) formalism \cite{Miller,Madison}. For highly energetic scattered electron, it is desirable to use the relativistic Dirac equation in the DW approximation (RDW method) \cite{Toshima,Srivastava,Sharma,Marucha}. 

In view of several applications of electron-atom scattering cross-sections such as in modelling metal vapour lasers and plasma plasma environments \cite{teubner}, learning insights into different physical processes in many natural and technological environments including the Earth's atmosphere 
and in the atmospheres of other planets and their satellites \cite{joshipura}, understanding electron-atom interactions \cite{Tawara} etc.. Theoretical studies on electron scattering by Be, Mg, Ca, Ne, Ar etc. atoms have been carried out earlier \cite{Jhanwar,Khare0,Khare,Baynard,ne1,ar1,Khoperskii,Kumar}. Most of these atoms have closed-shell electronic configurations. 

It is obvious from the above discussion that improvement in the accuracy of the scattering cross-section will depend on the accurate evaluation of atomic wave function and electric polarizabilities of the atom. Typical many-body methods employed to determine atomic wave functions are many-body perturbation theory (MBPT method), configuration interaction (CI) method, coupled-cluster (CC) method etc. among which CC method is treated as gold standard for its capability to incorporate electron correlations in the determination of atomic wave functions at a given approximation level \cite{bishop,szabo,crawford,bartlett}. Here, we employ CC method in the relativistic framework (RCC method) to evaluate the atomic wave functions. Though the (R)CC method has been applied earlier widely to calculate many spectroscopic properties to high accuracy, its capability to obtain the scattering cross-sections is not tested rigorously except in our first demonstrations in Mg$^+$ \cite{lalita} and Ca \cite{swati} to study scattering cross-sections in the plasma embedded and confined atom problems. Furthermore, atomic polarization effects on the scattering cross-sections are also quite significant. Often, contributions only from the electric dipole polarizabilities ($\alpha_d$) are considered in the construction of scattering potentials due to their dominant contributions. In recent calculations, it has been shown that contributions arising through the electric quadrupole ($\alpha_q$) and coupled dipole-quadrupole ($B$) polarizabilities are non-negligible \cite{Felipe,gribakin}. The aim of the present work is to provide general approaches to determine $\rho$, $\alpha_d$, $\alpha_q$ and $B$ values of atomic systems accurately by employing RCC method that can be used whenever required to obtain the elastic scattering cross-sections of an electron from the closed-shell atomic systems. For the representation purpose, we give the results for the Be, Ne, Mg and Ar atoms, but the scheme is very general and can be extended to atomic systems with open-shell configurations.

Apart from application of electric polarizabilities for determining electron scattering potentials, they are also immensely important for estimating Stark shifts of atomic energy levels. This is the reason why atomic polarizability studies are interesting on their own. In the literature, $\alpha_d$ has been studied extensively due to its predominant contribution to the energy shift, followed by $\alpha_q$ then $B$ in the presence of external electric field. Recently, we have presented linear response approach to determine the $\alpha_d$, $\alpha_q$ and $B$ values for Zn in the RCC and relativistic normal CC (RNCC) theory frameworks \cite{arup}. We had found that the results from the RCC and RNCC theories differ significantly in the commonly considered singles and doubles approximation. Here, we investigate $\rho$, $\alpha_d$, $\alpha_q$ and $B$ values from both the methods, and compare with the previously reported results, for the Be, Ne, Mg and Ar atoms. Using these values, we determine electron scattering potentials and show them by plotting against radial distances. Though these potentials are obtained by using the relativistic method, the estimated potentials can be adequately used both in the DW and RDW methods for calculating electron scattering cross-sections with different projectile energies. 

\section{Theory} \label{Model-Potential-theory}

For an spherically symmetric interaction potential $V(r)$ of the projectile electron with the target, the direct and exchange scattering amplitudes can be determined by \cite{newton}
\begin{eqnarray}
\label{direct}
f(k,\theta) &=& \frac{1}{2 \iota k} \sum_{l=0} ^\infty \left (
(l+1) (\exp({2\iota \delta_{\kappa = -l-1}}) -1) \right. \nonumber \\
&& + \left. l (\exp({2\iota \delta_{\kappa = l}}) -1) \right) P_l(\cos\theta)
\end{eqnarray} 
and
\begin{eqnarray}
\label{spin-flip}
g(k,\theta) &=& \frac{1}{2 \iota k} \sum_{l=0} ^\infty \left (
 \exp({2\iota \delta_{\kappa = l}})  \right. \nonumber \\
&& - \left.  \exp({2\iota \delta_{\kappa = -l-1}})  \right)  P_l^1(\cos\theta),
\end{eqnarray}
respectively. Here $k$ is the relativistic wave number, $\delta_{\kappa = -l-1, l}$ are the scattering phase shifts with  $\kappa = -l-1 $ and $\kappa = l$  referring to the relativistic quantum numbers for projectile electron with $ j = l+1/2$ and $j = l-1/2$, receptively. In above equation $\theta$ is the scattering angle, and $P_l(\cos\theta)$ and $P_l^1(\cos\theta)$ are Legendre polynomials and associated Legendre functions, respectively. Using these amplitudes, differential cross-sections per unit solid angle for spin unpolarized electrons can be calculated by
\begin{eqnarray}
\label{dcs}
\frac{d\sigma}{d\Omega} = |f(k,\theta)|^2 + |g(k,\theta)|^2 ,
\end{eqnarray}
from which integrated cross-sections can be estimated by integrating over the solid angle. In the (R)DW approximation, the first-order scattering amplitude of an electron from an atomic system with nuclear charge $Z$ and $N$ number of electrons can be expressed as 
\begin{eqnarray}
f(J_f,\mu_f;J_i,\mu_i,\theta) = 4 \pi^2 \sqrt{\frac{k_f}{k_i}} \langle  F_{DW}^{k_f} | H_{scat} |  F_{DW}^{k_i} \rangle ,
\end{eqnarray}
where $J$ and $\mu$ represent for angular momenta of the states of atomic target and scattered electron respectively, $k$ is the momentum of the scattered electron and $F_{DF}$ are the (R)DW wave functions while the subscripts $i$ denotes for initial state and $f$ denotes for final state.
A similar expression can be given for $g$. In the DW method, the effective scattering Hamiltonian in atomic units (a.u.) is given by  
\begin{eqnarray}
H_{scat} = -\frac{1}{2} \nabla^2 + V(r)
\end{eqnarray}
whereas in the RDW method, it is given by
\begin{eqnarray}
H_{scat} = c \alpha \cdot \mathbf{p}+ \beta c^{2} + V(r) .
\end{eqnarray}
Here $c$ is the speed of light, $\alpha$ and $\beta$ are the Dirac matrices and $V(r)$ is the scattering potential. For accurate determination of scattering cross-sections, it is imperative to obtain $V(r)$ accurately. In a more convenient form, $V(r)$ can be expressed as \cite{connel}
\begin{eqnarray}
V(r) = V_{st}(r) + V_{ex}(r) + V_{pol}(r) ,
\end{eqnarray}
where $V_{st}(r)$, $V_{ex}(r)$ and $V_{pol}(r)$ are known as the static, exchange and polarization potentials, respectively. The static potential can have contributions from nuclear potential ($V_{nuc}(r)$) and direct electron-electron Coulomb interaction potential $V_{C}(r)$; i.e. $V_{st}(r)=V_{nuc}(r)+V_C(r)$. Usually, a point-like atomic nucleus is considered in the scattering cross-section calculations 
by defining $V_{nuc}(r)=-\frac{Z}{r}$ for the atomic number of the system $Z$. In the present work, we have used Fermi-charge distribution, given by
is given by
\begin{eqnarray}
\rho_A(r) = \frac{\rho_0}{1+e^{(r-c)/a}} ,
\end{eqnarray}
where $\rho_0$ is the normalization constant, $c$ is the half-charge radius and $a=2.3/4ln(3)$ is known as the skin thickness, to take into 
account of finite size effect of the nucleus. This corresponds to expression for the nuclear potential as \cite{Estevez}
\begin{eqnarray}
 V_{nuc}(r) = -\frac{Z}{\mathcal{N}r} \times \ \ \ \ \ \ \ \ \ \ \ \ \ \ \ \ \ \ \ \ \ \ \ \ \ \ \ \ \ \ \ \ \ \ \ \ \ \ \ \  \nonumber\\
\left\{\begin{array}{rl}
\frac{1}{c}(\frac{3}{2}+\frac{a^2\pi^2}{2c^2}-\frac{r^2}{2c^2}+\frac{3a^2}{c^2}P_2^+\frac{6a^3}{c^2r}(S_3-P_3^+)) & \mbox{for $r_i \leq c$}\\
\frac{1}{r_i}(1+\frac{a62\pi^2}{c^2}-\frac{3a^2r}{c^3}P_2^-+\frac{6a^3}{c^3}(S_3-P_3^-))                           & \mbox{for $r_i >c$} ,
\end{array}\right.   
\label{eq12}
\end{eqnarray}
where the factors are 
\begin{eqnarray}
\mathcal{N} &=& 1+ \frac{a^2\pi^2}{c^2} + \frac{6a^3}{c^3}S_3  \nonumber \\
\text{with} \ \ \ \ S_k &=& \sum_{l=1}^{\infty} \frac{(-1)^{l-1}}{l^k}e^{-lc/a} \ \ \  \nonumber \\
\text{and} \ \ \ \ P_k^{\pm} &=& \sum_{l=1}^{\infty} \frac{(-1)^{l-1}}{l^k}e^{\pm l(r-c)/a} . 
\end{eqnarray}

Similarly, we can express $V_C(r) = \sum_{b=1}^{N_e} \langle \phi_b |  \frac{1}{|{\vec r}-{\vec r}_b|} | \phi_b \rangle $ with $N_e$ denoting total number 
of electrons of the target atom and $| \phi_b \rangle$ is the single particle wave function of the atomic electron $b$ such that
\begin{eqnarray}
\frac{1 }{|{\vec r}_i-{\vec r}_j|} &=& \sum_{k=0}^{\infty} \frac{4\pi}{2k+1} \frac{r_{<}^k}{r_{>}^{k+1}} 
       \sum_{q=-k}^k Y_{q}^{k\ast}(\theta,\varphi)Y^k_{q}(\theta,\varphi) , \ \ \ \
\label{veff}
\end{eqnarray}
where  $r_> =$max($r_i,r_j$), $r_<=$ min($r_i,r_j$), and $Y^k_{q}(\theta,\varphi)$ is the spherical harmonics of rank $k$ with its component $q$. 
In terms of the Racah operator ($C^k_{q}$), the above expression is given by a scalar product as
\begin{eqnarray}
\frac{1 }{|{\vec r}_i-{\vec r}_j|} &=& \sum_{k=0}^{\infty} \frac{r_{<}^k}{r_{>}^{k+1}} \mbox{\boldmath$\rm C$}^k(\hat{r}_i) \cdot \mbox{\boldmath$\rm C$}^k(\hat{r}_j).
\end{eqnarray}

In the Dirac theory, the single particle orbital wave functions are given by
\begin{equation}
\vert \phi (r) \rangle = \frac{1}{r} \begin{pmatrix} P(r)\chi_{j m_j l_L} (\theta,\varphi) \\ i Q(r) \chi_{j m_j l_S}(\theta,\varphi) \end{pmatrix} ,
\end{equation}
where the upper and lower components are the large and small components of the single particle wave function, respectively, $P(r)$ and $Q(r)$
denote the radial parts of these components, and the $\chi$'s denote the spin angular parts of each component which depend on the quantum numbers
$j$, $m_j$, and $l$. $l_L$ denotes $l$ for the large component, while $l_S$ denotes $l$ for the small component. Thus for a closed-shell atomic 
target like the ones that are under consideration in this work, we can have 
\begin{eqnarray}
 V_C(r) = \sum_{b} (2j_b+1) \int_0^{\infty} dr_b \frac{1}{r_{>}} \left [ P_b^2 (r_b) +  Q_b^2 (r_b) \right ].
\end{eqnarray}
It is worth noting that for open-shell atomic targets, there will be finite value of multipoles $k$ in the above expression and the computation of $V_C(r)$ will be slightly difficult but it is possible \cite{lalita}. Using density function formalism, the above expression can be given by
\begin{eqnarray} 
\label{vstatic}
V_{C}(r) = \sum_b \left [ \frac{1}{r}  \int_0^r dr_b \rho_b(r_b) r_b'^2  + \int_r^\infty dr_b \rho_b(r_b) r_b  \right ] ,
\end{eqnarray}
where the atomic density function is given by $\hat \rho (r) = \sum_i \hat \rho_i (r) = \sum_i |\phi_i \rangle \langle \phi_i|$ with 
\begin{eqnarray}
 \langle \phi_j | \hat \rho_i (r) | \phi_k \rangle =  \delta_{ji} \delta_{ik} \left ( P_j(r) P_k (r)+ Q_j(r) Q_k(r) \right ) .
\end{eqnarray}

It is not possible to determine $V_{ex}(r)$ separately as it depends on the wave function of the scattered electron itself. However, it can be 
approximately estimated by using the Hara free electron gas model, given by \cite{Hara}
\begin{eqnarray}
 V_{ex}(r) = - \frac{2}{\pi} K_F(r) F[\eta(r)] ,
\end{eqnarray}
where the Fermi momentum $K_F(r) = (3 \pi \rho(r) )^{1/3}$ and $F(\eta) = \frac{1}{2} + \frac{1-\eta^2}{4 \eta} ln \left \vert \frac{1+ \eta}{1-\eta} \right \vert$
with $\eta(r) = \frac{K(r)}{K_F(r)}$ for the local electron momentum given by
\begin{eqnarray}
 K^2(r) = K_F^2 + 2 I+ k^2 .
\end{eqnarray}
Here $I$ denotes for ionization potential (IP) of the target atom and $k^2/2$ is the kinetic energy of the projectile electron. It means that evaluation of $V_{ex}(r)$ requires atomic density function and IP of the atom along with the kinetic energy of the projectile. Since 
the kinetic energy of the projectile is arbitrary, we provide here only the $\rho(r)$ values while IPs can be used from the experimental data. 

The polarization potential is given by \cite{Burrow, Arretche}
\begin{eqnarray}
 V_{pol}(r) &=& - \left ( \frac{\alpha_d}{2r^4} +  \frac{\alpha_q}{2r^6} - \frac{B}{2r^7} + O(1/r^8) \right ) \nonumber \\ && \times \left [ 1- e^{(r/r_c)^6} \right ], 
\end{eqnarray}
where $\alpha_d$, $\alpha_q$ and $B$ are known as second-order dipole, second-order quadrupole and third-order dipole-quadrupole polarizabilities respectively. $O(1/r^8)$ corresponds to higher-order polarizability contributions and neglected here. $r_c$ is a adjustable parameter, which can be determined by estimating IP using the above potential in the equation-of-motion, is supposed to be different for different atoms and also for different level of approximation in the above expression. For convenience and demonstration purpose, without losing much accuracy, we have considered $r_c= 3.5$ in atomic units (a.u.) for all the considered atoms \cite{Burrow}.

In the following section, we present the RCC method to estimate $\rho(r)$, $V_{st}$, $\alpha_d$, $\alpha_q$ and $B$ in the closed-shell atomic systems. In place of calculating $V_{st}(r)$ directly using the RCC theory, we estimate it by evaluating $V_{nuc}(r)$ and $V_C(r)$ separately in which $V_C(r)$ is obtained from the $\rho(r)$ values. Again, the expectation values of the operators are evaluated using the standard RCC and RNCC theory frameworks, and the results are compared with the earlier reported literature values.

\begin{figure*}[t] 
\includegraphics[height=6cm,width=8.5cm]{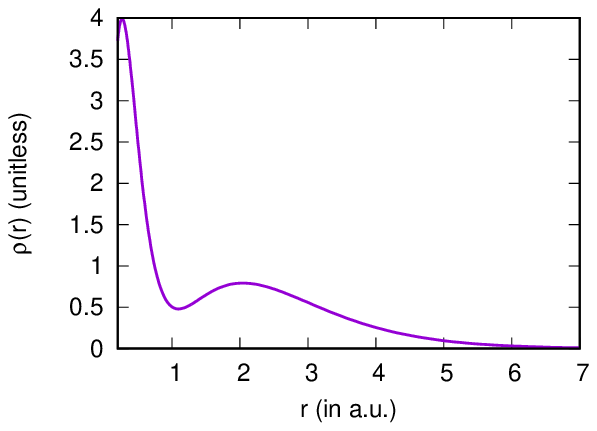}
\includegraphics[height=6cm,width=8.5cm]{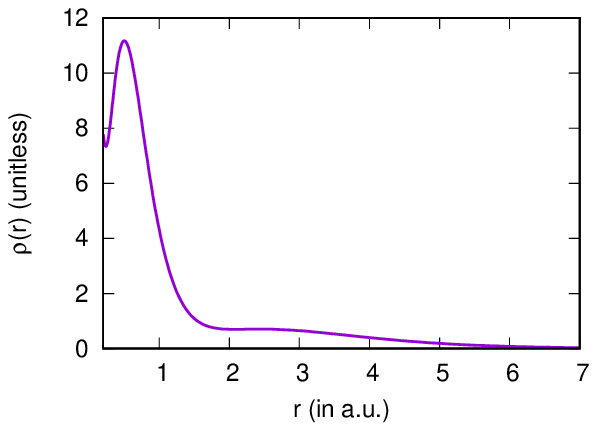}
\includegraphics[height=6cm,width=8.5cm]{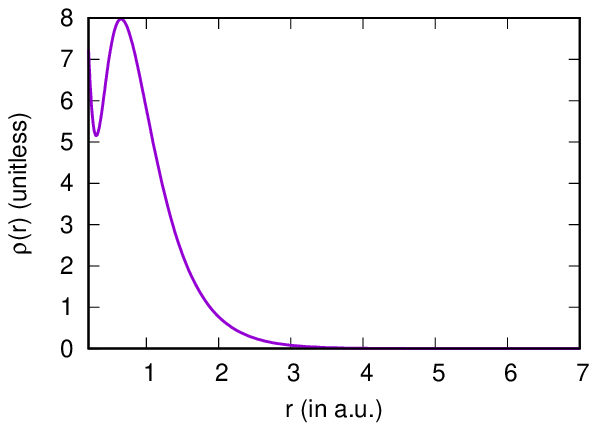}
\includegraphics[height=6cm,width=8.5cm]{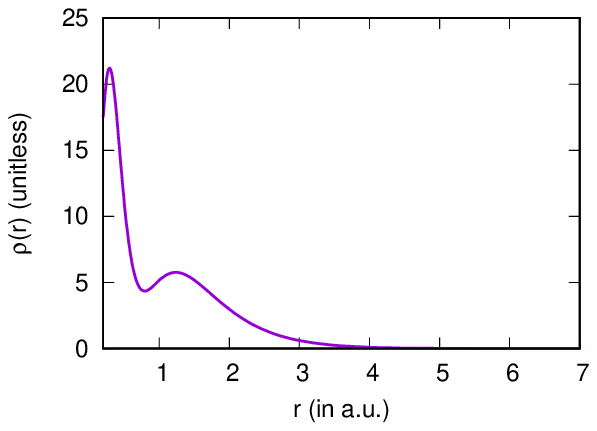}
\caption{Density profiles of the (a) Be, (b) Mg, (c) Ne and (d) Ar atoms obtained using the DHF method in their ground states. The radial distances (r) are given in atomic units (a.u.) while density $\rho$(r) are unitless.}
\label{densfig}
\end{figure*}

\begin{figure*}[t] 
\includegraphics[height=6cm,width=8.5cm]{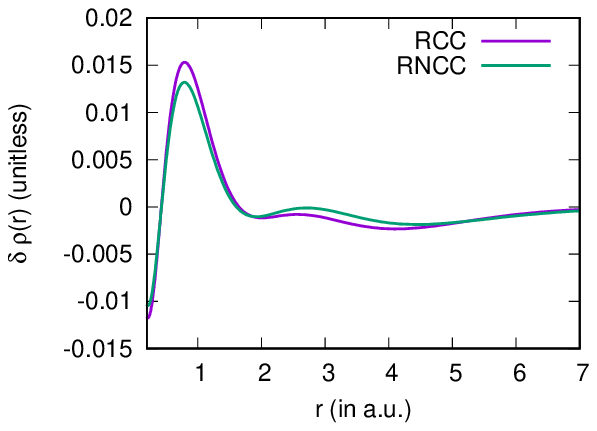}
\includegraphics[height=6cm,width=8.5cm]{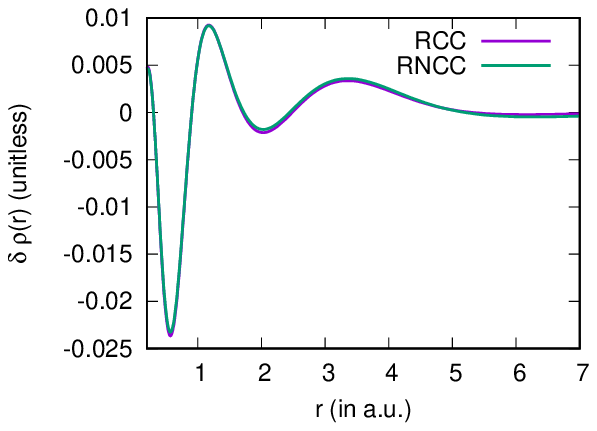}
\includegraphics[height=6cm,width=8.5cm]{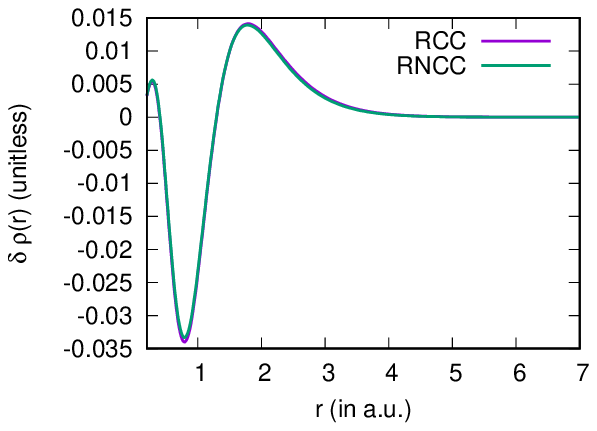}
\includegraphics[height=6cm,width=8.5cm]{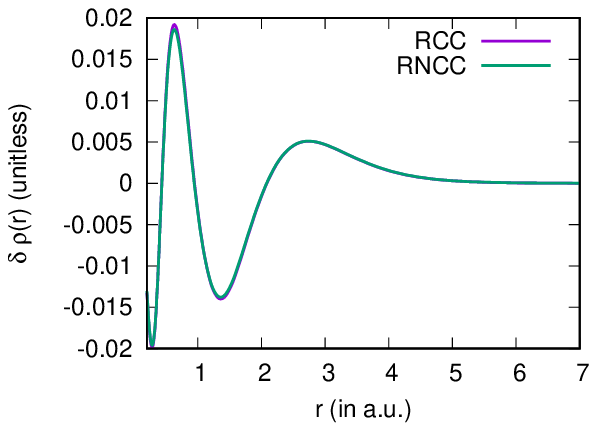}
\caption{Correlation contributions to $\rho(r)$ (shown in figure as $\delta \rho(r)$) from the RCC and RNCC methods in (a) Be, (b) Mg, (c) Ne and (d) Ar. As seen, the $\delta \rho(r)$ values are coming out to be almost same through the RCC and RNCC methods in all atoms except in Be; in which slight differences are noticed.}
\label{densdiff}
\end{figure*}

\section{Methods for Calculations}

Since $\alpha_d$, $\alpha_q$ and $B$ are determined by treating electric dipole operator $D$ and quadrupole operator $Q$ as external perturbation, atomic wave functions without these external operators are denoted with superscript $0$ ($| \Psi_0^{(0)} \rangle $). We have considered Dirac-Coulomb Hamiltonian to determine these unperturbed wave functions, given by
\begin{equation}
H_0=\sum_{i=1}^{N_e}\left[c \alpha_{i} \cdot \mathbf{p}_{i}+\left(\beta_{i}-1\right) c^{2}+V_{n u c}\left(r_{i}\right)+\sum_{j>i} \frac{1}{r_{i j}}\right] ,
\end{equation}
where $r_{ij}=|\vec{r}_i - \vec{r}_j|$ is the inter-electronic separation between the electrons located at the $r_i$ and $r_j$ radial positions with respect to the center of the nucleus.   

The density matrix of the atomic state $| \Psi_0^{(0)} \rangle$ can be determined by
\begin{eqnarray}
\rho(r) &=&  \frac{\langle \Psi_0^{(0)}| \hat \rho(r) | \Psi_0^{(0)} \rangle }{\langle \Psi_0^{(0)}|\Psi_0^{(0)} \rangle } .
\end{eqnarray}

Following Ref. \cite{arup}, the expressions for $\alpha_d$, $\alpha_q$ and $B$ of the ground state of a closed-shell system can be given by 
\begin{eqnarray}
\alpha_d &=& 2 \frac{\langle \Psi_0^{(0)}|D| \Psi_0^{(d,1)} \rangle }{\langle \Psi_0^{(0)}|\Psi_0^{(0)} \rangle } , \nonumber \\
\alpha_q &=& 2 \frac{\langle \Psi_0^{(0)}|Q| \Psi_0^{(q,1)} \rangle }{\langle \Psi_0^{(0)}|\Psi_0^{(0)} \rangle }
\end{eqnarray}
and
\begin{eqnarray}
B &=& 2 \frac{\langle \Psi_0^{(d,1)}|D| \Psi_0^{(q,1)} \rangle }{\langle \Psi_0^{(0)}|\Psi_0^{(0)} \rangle } ,
\end{eqnarray}
where $|\Psi_0^{(0)} \rangle$ and $|\Psi_0^{(o,1)} \rangle$ are the zeroth-order wave function and the first-order wave function of the atom due to an operator $O\equiv D$ or $Q$. 

It is obvious from the above expressions that accurate evaluations of $\alpha_d$, $\alpha_q$ and $B$ depend on the many-body method employed 
to determine $|\Psi_0^{(0)} \rangle $ and $|\Psi_0^{(o,1)} \rangle$. These wave functions can be determined by solving the following equations
\begin{eqnarray}
H_0  |\Psi_0^{(0)} \rangle = E_0^{(0)} |\Psi_0^{(0)} \rangle
\end{eqnarray}
and 
\begin{eqnarray}
(H_0 - E_0^{(0)}) |\Psi_0^{(o,1)} \rangle = (E_0^{(o,1)} - O) |\Psi_0^{(0)} \rangle 
\end{eqnarray}
with the first-order energy correction $E_0^{(o,1)}$ due to $O$, which is zero in the present study.

Our intention here is to demonstrate evaluation of $\rho(r)$, $\alpha_d$, $\alpha_q$ and $B$ in the closed-shell atoms using the RCC and RNCC theories that can be used for constructing the electron-atom scattering potentials. In the RCC theory, we can express  \cite{bijaya,yashpal}
\begin{equation}
|\Psi_0^{(0)} \rangle=e^{T^{(0)}} |\Phi_0 \rangle , \label{eqcc1}
\end{equation}
and
\begin{equation}
|\Psi_{0}^{(o,1)} \rangle = e^{T^{(0)}} T^{(o,1)} |\Phi_0 \rangle , \label{eqcc2}
\end{equation}
where $T^{(0)}$ accounts for electron correlation effects, and $T^{(o,1)}$ includes electron correlations along with the effect due to $O$ while acting on the Dirac-Hartree-Fock (DHF) wave function $|\Phi_0 \rangle$ of the system.

In this approach, the expressions for $\rho$,  $\alpha_d$, $\alpha_q$ and $B$ are given by \cite{arup}
\begin{eqnarray}
\rho(r) &=&  \frac{\langle \Phi_0 | e^{T^{(0)\dagger}} \hat \rho (r) e^{T^{(0)}}  | \Phi_0 \rangle} {\langle \Phi_0 | e^{T^{(0)\dagger}}  e^{T^{(0)}} | \Phi_0 \rangle} , \\
\alpha_d &=& 2 \frac{\langle \Phi_0 | e^{T^{(0)\dagger}} D  e^{T^{(0)}} T^{(d,1)} | \Phi_0 \rangle} {\langle \Phi_0 | e^{T^{(0)\dagger}}  e^{T^{(0)}} | \Phi_0 \rangle} , \\
\alpha_q &=& 2 \frac{\langle \Phi_0 | e^{T^{(0)\dagger}} Q  e^{T^{(0)}} T^{(q,1)} | \Phi_0 \rangle} {\langle \Phi_0 | e^{T^{(0)\dagger}}  e^{T^{(0)}} | \Phi_0 \rangle}
\end{eqnarray}
and
\begin{eqnarray}
B &=& 2 \frac{\langle \Phi_0 |T^{(d,1)\dagger}e^{T^{(0)\dagger}}  D e^{T^{(0)}} T^{(q,1)}| \Phi_0 \rangle} {\langle \Phi_0 | e^{T^{(0)\dagger}}  e^{T^{(0)}} | \Phi_0 \rangle} .
\end{eqnarray}

Evaluating the above expressions involve two major challenges, even after making approximations in the level of excitations in the RCC calculations. The first being that it has two non-terminating series in the numerator and denominator. The second that the numerator can have factors both connected and disconnected with the operators $D$ or $Q$. These problems can be partially addressed by defining normal-order form of operators with respect to $|\Phi_0 \rangle$, in which the above expressions can be simplified to \cite{yashpal1,prasanna}
\begin{eqnarray}
\rho(r) &=&  \langle \Phi_0 | e^{T^{(0)\dagger}} \hat \rho(r)  e^{T^{(0)}} T^{(d,1)} | \Phi_0 \rangle_c ,  \\
\alpha_d &=& 2 \langle \Phi_0 | e^{T^{(0)\dagger}} D  e^{T^{(0)}} T^{(d,1)} | \Phi_0 \rangle_c ,  \\
\alpha_q &=& 2 \langle \Phi_0 | e^{T^{(0)\dagger}} Q  e^{T^{(0)}} T^{(q,1)} | \Phi_0 \rangle_c
\end{eqnarray}
and
\begin{eqnarray}
B &= & 2 \langle \Phi_0 |T^{(d,1)\dagger}e^{T^{(0)\dagger}}  D e^{T^{(0)}} T^{(q,1)}| \Phi_0 \rangle_c ,
\end{eqnarray}
where subscript $c$ denotes connected terms only appearing within the respective expression. Though it removed the non-terminating series appearing in the denominator, it still contains a non-terminating series in the numerator. Further, the above expressions with connected terms hold good only when there is no approximation made in the $T$ operator. In practice, $T$ is truncated like our RCCSD method. Again, these expressions do not satisfies the Hellman-Feynman theorem \cite{bishop}. All these problems can be circumvented by the RNCC theory.

\begin{table}[t]
\caption{Our calculated values of $\alpha_d$, $\alpha_q$ and $B$ (in a.u.) of the Be, Mg, Ne and Ar atoms from the DHF, RCCSD and RNCSSD methods. These values are also compared with the available precise values from the literature.}
\begin{tabular}{lcccc}
\hline \hline
Property & DHF & RCCSD & RNCCSD & Others \\
\hline \\
\multicolumn{5}{c}{\underline{Be atom}} \\
 $\alpha_d$   & 30.53 & 38.33 & 37.40 & 37.739(30) \cite{thakkar} \\
              &       &       &       &   37.76(22) \cite{porsev} \\
              &       &       &   & 37.86(17) \cite{yashpal2}\\
             &       &       &   & 37.74(3) \cite{poldata}\\  
 $\alpha_q$   & 220.15 & 299.82 & 304.34 & 300.96 \cite{thakkar} \\
             &       &       &       &   300.6(3) \cite{porsev} \\
  $B$         & $-1218.38$  & $-2729.17$ & $-2172.95$ & $-2100(60)$ \cite{thakkar} \\
 \hline \\
  \multicolumn{5}{c}{\underline{Mg atom}} \\
 $\alpha_d$   & 54.94 & 71.74 & 69.40 &  71.22(36) \cite{thakkar} \\
            &       &       &       &   71.3(7) \cite{porsev} \\
           &       &       &   & 72.54(50) \cite{yashpal2}\\
  &       &       &   & 71.2(4) \cite{poldata}\\          
 $\alpha_q$   & 567.37 & 809.56 & 797.91 & 813.9(16.3) \cite{thakkar} \\
          &       &       &       &   812(6) \cite{porsev} \\
  $B$         & $-3847.89$  & $-9293.74$ & $-7226.24$ & $-7750(780)$ \cite{thakkar} \\
  \hline  \\
 \multicolumn{5}{c}{\underline{Ne atom}} \\
 $\alpha_d$   & 1.98 & 2.70 & 2.62 &  2.6669(8) \cite{expt1} \\
              &      &      &      & 2.652(15) \cite{yashpal2} \\
       &       &       &   & 2.66110(3) \cite{poldata}\\   
 &       &       &   & 2.64 \cite{taylor}\\      
 $\alpha_q$   & 4.76 & 7.48 & 7.09 & 7.52(15) \cite{thakkar} \\
  &       &       &   & 7.36 \cite{taylor}\\ 
  $B$         & $-6.15$  & $-14.38$ & $-11.67$ &  $-18.12(54)$ \cite{thakkar} \\
  &       &       &   & $-17.27$ \cite{taylor}\\ 
  \hline \\
  \multicolumn{5}{c}{\underline{Ar atom}} \\
 $\alpha_d$   & 10.15 & 11.21 & 11.15 & 11.083(7) \cite{expt2} \\
  &       &       &   & 11.070(7) \cite{expt3} \\
      &      &      &      & 11.089(4) \cite{yashpal2} \\
         &       &       &   & 11.083(7) \cite{poldata}\\
       &       &       &   & 11.33 \cite{ivan}\\\
      &       &       &   & 10.73 \cite{thar} \\
 $\alpha_q$   & 37.19 & 51.61 & 50.33 &  53.37(1.07)  \cite{thakkar} \\
       &       &       &   & 53.22 \cite{ivan}\\
        &       &       &   & 49.46 \cite{thar} \\
 $B$  & $-71.07$ & $-140.53$ & $-115.35$ &  $-159(8)$ \cite{thakkar} \\
       &       &       &   & $-167.5$ \cite{ivan}\\
  &       &       &   & $-141$ \cite{thar} \\
 \hline \hline
\end{tabular}
\label{tab1}
\end{table}

In the RNCC theory, the ket state is same as the RCC theory but the bra state is replaced by
\begin{eqnarray}
 \langle \tilde{\Psi}^{(0)} | = \langle \Phi_0 | (1+\Lambda^{(0)}) e^{-T^{(0)}} ,
\end{eqnarray}
with a de-excitation operator $\Lambda^{(0)}$ that satisfies
\begin{eqnarray}
 \langle \tilde{\Psi}^{(0)} | \Psi^{(0)} \rangle = \langle \Phi_0 | (1+ \Lambda^{(0)}) e^{-T^{(0)}} e^{T^{(0)}} |\Phi_0 \rangle =1 .
\end{eqnarray}
It can be shown that eigenvalues of both $\langle \Psi^{(0)} |$ and $\langle \tilde{\Psi}^{(0)}|$ are same if 
\begin{eqnarray}
\langle \Phi_0 |\Lambda \bar{ H}_0 |\Phi_0 \rangle= 0 ,
\end{eqnarray}
where $\bar{H} = e^{- T^{(0)}} H_0 e^{ T^{(0)}} = (He^T)_c$. 

Now, we can write the first-order perturbed wave function in the RNCC theory as \cite{arup,bijaya2}
\begin{eqnarray}
 \langle \tilde{\Psi}^{(o,1)} | = \langle \Phi_0 | \left [ \Lambda^{(o,1)} + (1+\Lambda^{(0)}) T^{(o,1)} \right ] e^{-T^{(0)}} .
\end{eqnarray}
Consequently, the RNCC expressions for $\rho(r)$, $\alpha_d$, $\alpha_q$ and $B$ are given by
\begin{eqnarray}\label{polnrcc}
\rho(r) &=& \langle\Phi_0 |\left(1+\Lambda^{(0)}\right)  \tilde{{\hat \rho}}(r)  | \Phi_0 \rangle , \\
\alpha_d &=& \langle\Phi_0 |\left(1+\Lambda^{(0)}\right) \tilde{D} T^{(d,1)}+\Lambda^{(d,1)} \tilde{D} | \Phi_0 \rangle , \label{eqnd} \\
\alpha_q &=& \langle\Phi_0 |\left(1+\Lambda^{(0)}\right) \tilde{Q} T^{(q,1)}+\Lambda^{(q,1)} \tilde{Q} | \Phi_0 \rangle \label{eqnq}
\end{eqnarray}
and
\begin{eqnarray}
B &=& \langle\Phi_0 |\Lambda^{(d,1)}  \tilde{D} T^{(q,1)} + \Lambda^{(q,1)}  \tilde{D} T^{(d,1)}| \Phi_0 \rangle , \label{eqnb}
\end{eqnarray}
where $\tilde{O}= (Oe^{T^{(0)}})_c$. In the RNCC theory, we also consider only the singles and doubles excitations (RNCCSD method) to carry out the
calculations. It is worth mentioning here that the next leading-order electron correlation effects to $\rho(r)$, $\alpha_d$, $\alpha_q$ and $B$
arising through the higher-level excitations will converge faster in the RNCC theory than the RCC theory \cite{arup,bijaya2}.

\begin{figure*}[t] 
\includegraphics[height=6cm,width=8.5cm]{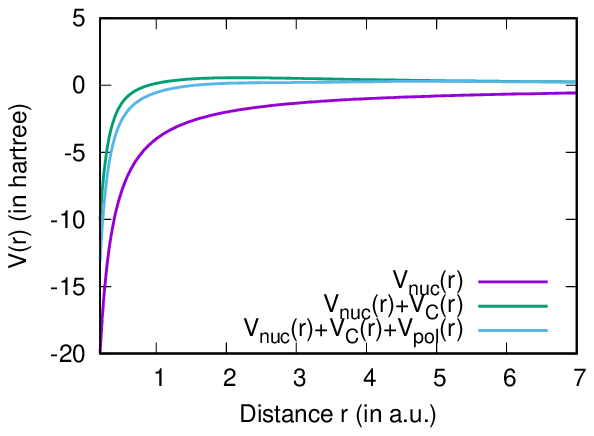}
\includegraphics[height=6cm,width=8.5cm]{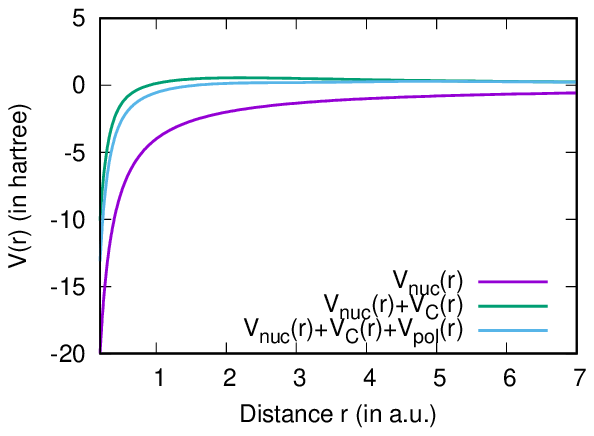}
\includegraphics[height=6cm,width=8.5cm]{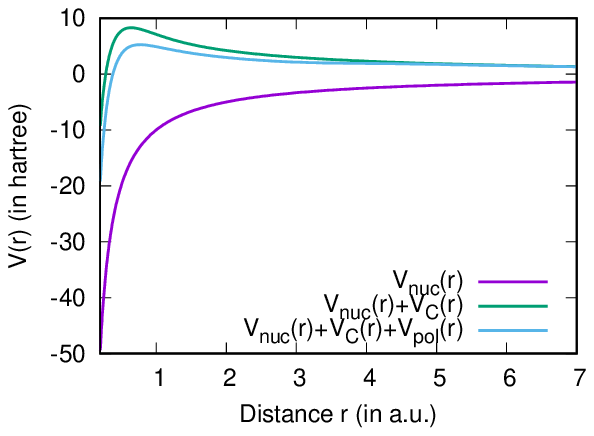}
\includegraphics[height=6cm,width=8.5cm]{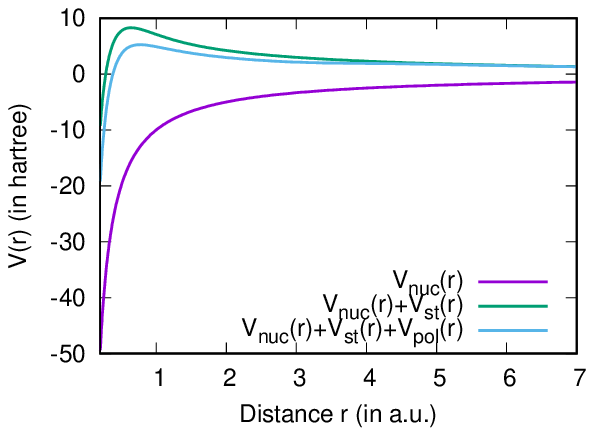}
\caption{Plots demonstrating comparative analyses of contributions from $V_{nuc}(r)$, $V_C(r)$ and $V_{pol}(r)$ to the electron scattering potential $V(r)$ from the Be and Mg alkaline-earth atoms. In Figs. (a) and (b), results are given from the RCCSD and RNCCSD methods respectively for the Be atom. Results from the RCCSD and RNCCSD methods are shown in Figs. (c) and (d) respectively for the Mg atom. All quantities are given in a.u..}
\label{potbemg}
\end{figure*}

\section{Results and Discussion}\label{RnD}

We first evaluate density functions $\rho(r)$ of the ground states of the Be, Mg, Ne and Ar atoms. Since correlation contributions, i.e. differences between the DHF and RCC/RNCC values (given as $\delta \rho(r)$), to these functions are very small compared to the DHF values, we give these contributions separately. In Fig. \ref{densfig}, we plot $\rho(r)$ values from the DHF method, while correlation contributions $\delta \rho(r)$ from the RCCSD and RNCCSD methods are shown in Fig. \ref{densdiff}. As can be seen from the first figure, the density profile of Be, Mg, Ne and Ar look differently. This suggests that the electronic charge distribution among these atoms are quite different. From the second figure, we see that there is slight differences in the correlation contributions from the RCCSD and RNCCSD methods in Be, while in other atoms there are not much differences observed. As mentioned earlier, accurate values of $\alpha_d$, $\alpha_q$ and $B$ are important in determining $V_{pol}(r)$ for the electron-atom scattering problem. Therefore, roles of electron correlation effects through the RCCSD and RNCCSD methods in the above atoms can be understood better through the calculations of electric polarizabilities.

To our knowledge, there are no calculations of $\rho(r)$ of the considered atoms available explicitly by the (R)CC methods earlier. In a recent work \cite{Corzo}, a CI method was employed in the non-relativistic framework to determine density functions for studying quantum potential neural network of the lithium, Be and Ne atoms. We find that our density function behaviors in Be and Ne are almost matching with the density functions of these atoms reported in Ref. \cite{Corzo}. We could not find any reference reporting density functions of Mg and Ar directly, however from the analyses of radial function distributions in Ne and Ar shown in Ref. \cite{Zaklika} we assume that the behavior of the density functions of the Ar atom obtained by us using the DHF method follow the correct trend. Also in a different work \cite{Diouf}, calculations of the $\rho(r)$ values in carbon atom follow similar trends like our results for Mg. From all these analyses, we presume that our $\rho(r)$ values for Mg should be correct. Since the previous works do not discuss about $\delta \rho(r)$ contributions explicitly, we are unable to compare our findings on these values with any other calculations. 
  
In Table \ref{tab1}, we present the $\alpha_d$, $\alpha_q$ and $B$ values calculated using the DHF, RCCSD and RNCCSD methods. It can be seen from this table that there are large differences between the results from the DHF and RCCSD methods. These differences become larger in the determination of $\alpha_q$ followed by the $B$ values. The RCCSD values of $B$ in the alkaline-earth atoms are about 2.5 times larger than the DHF values. In all atoms, the RNCCSD values of $\alpha_d$, $\alpha_q$ and $B$ are seen to be lower than the RCCSD values except in the determination of $\alpha_q$ in the Be atom. The $\alpha_d$ values from the RCCSD and RNCCSD methods are almost close to each other, but there are significant differences seen among the $\alpha_q$ values of the RCCSD and RNCCSD methods. These differences are quite prominent in the evaluation of the $B$ values. As discussed in the previous section, an approximated RNCC method is more reliable in the determination of properties than an approximated RCC method, so we believe that our RNCCSD results are more accurate and should be treated in this work as our final results.

\begin{figure*}[t] 
\includegraphics[height=6cm,width=8.5cm]{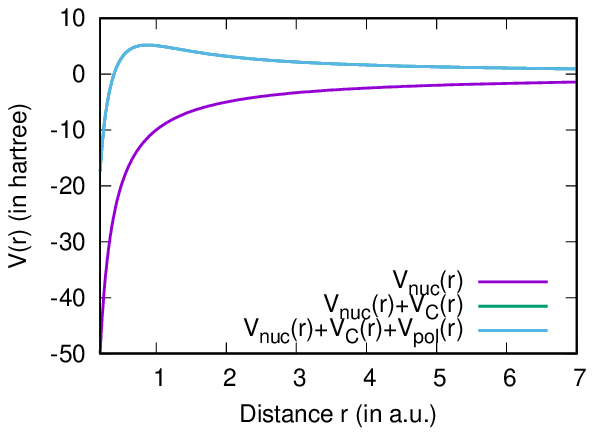}
\includegraphics[height=6cm,width=8.5cm]{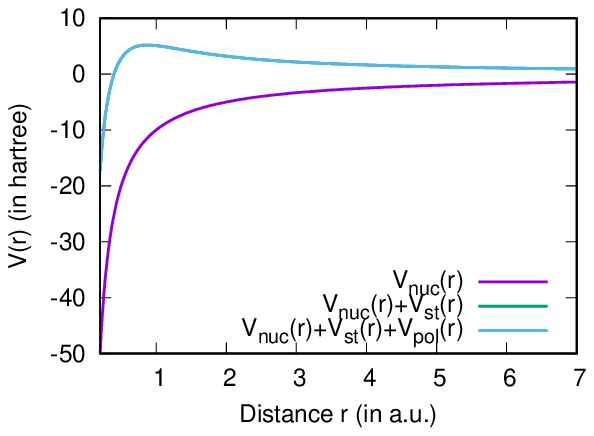}
\includegraphics[height=6cm,width=8.5cm]{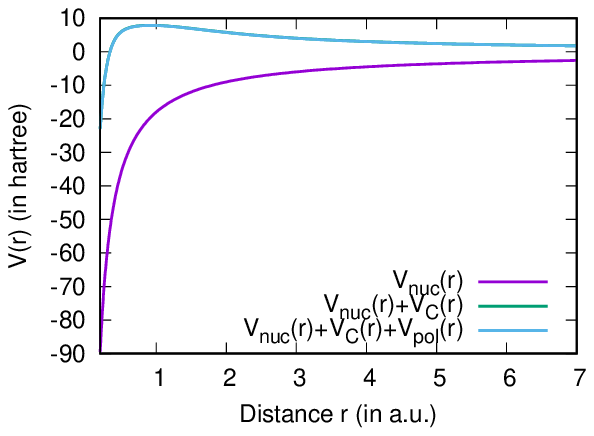}
\includegraphics[height=6cm,width=8.5cm]{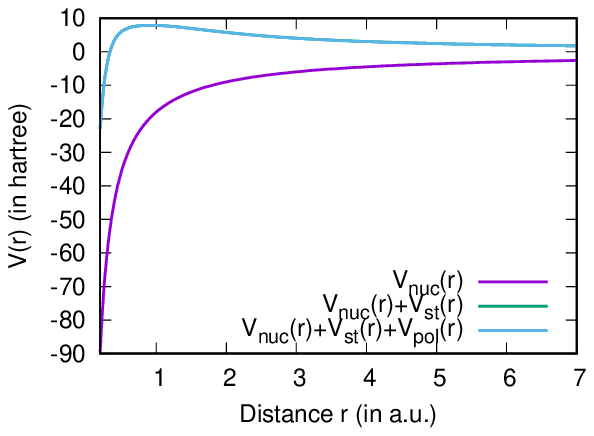}
\caption{Plots demonstrating different contributions to $V(r)$ from the Ne and Ar noble atoms. In Figs. (a) and (b), results are given from the RCCSD and RNCCSD methods respectively for the Ne atom, while these results from the RCCSD and RNCCSD methods for the Ar atom are shown in Figs. (c) and (d) respectively. All quantities are given in a.u..}
\label{potnear}
\end{figure*}

Due to immense applications of electric polarizabilities in various experimental applications, a number of theoretical calculations are carried out in literature.
We consider previous results from the experiments \cite{expt1,expt2,expt3}, references that provide compilation of earlier data \cite{thakkar,poldata}, from our previous RCC calculations \cite{yashpal2} and from the papers that report most of these quantities using a single many-body method \cite{porsev,ivan,thar,taylor}. Results other publications are mostly summarized in Refs. \cite{thakkar,poldata} in detail. Many of the earlier theoretical studies have determined the $\alpha_d$ values, where less theoretical results can be found for $\alpha_q$ of the considered atoms. To our knowledge, only a few non-relativistic calculations for the $B$ values of the considered Be, Mg, Ne and Ar atoms have been reported \cite{thakkar,ivan,taylor,thar}. Also, we did not find any experimental results of $\alpha_d$ for Be and Mg, but precise measured $\alpha_d$ values are available for Ne and Ar.  Our both the RCCSD and RNCSSD values are in agreement with the previous calculations. We also observe that our RCCSD value of $\alpha_q$ is more close with the previously reported precise calculation than the RNCCSD value, but this trend is different for the $\alpha_d$ and $B$ values. These findings are slightly different for the Mg atom, where we see that both the $\alpha_d$ and $\alpha_q$ values from our RCCSD method match well with the previously reported accurate calculations but the RNCCSD value for $B$ agrees better with the previous calculation \cite{thakkar}. From these comparisons it is not possible to argue that RCCSD method values are more accurate over the RNCCSD results wherever they agree with the previous calculations unless they are verified by the experiments. Again, the $\alpha_q$ and $B$ values were estimated using the finite-field (FF) approach earlier which are not numerically reliably. The experimental value of $\alpha_d$ value in Ne is very precise, and comparison of theoretical results with this value can indicate validity of our employed many-body methods. We have also compared our RNCCSD values of $\alpha_d$ and $\alpha_q$ with the literature values in Table \ref{tab1}. In Be, there are several calculations of $\alpha_d$ available and we have listed some of the precise theoretical results in the above table from the CC and RCC calculations. A few calculations of $\alpha_q$ of the considered atoms including Be have been reported using non-relativistic variation-perturbation methods in the finite-field (FF) approach \cite{thakkar,ivan,thar,taylor} and using combined CI and MBPT (CI$+$MBPT) method in the sum-over-states approach \cite{porsev}. Our RNCCSD $\alpha_d$ value matches with the previously estimated values. We found a slight difference for the $\alpha_q$ value from the RNCCSD method and previously reported precise value using the CI$+$MBPT method \cite{porsev}. Our RCCSD value of $\alpha_d$ in Mg agrees well with the previously calculated values using various many-body methods. However, it can be noticed that the previously reported $\alpha_d$ values from different calculations spread over a wide-range. This is owing to the large electron correlation effects exhibited by both the valence electrons of the Mg atom. Nonetheless, our RNCCSD value of $\alpha_d$ is also close to other calculations. However, we find that our RCCSD value for $\alpha_q$ is more close with the previous calculation while the RNCCSD result differs significantly from the earlier calculations. We cannot say with confidence from this difference that the RCCSD value is more accurate than the RNCCSD result. This is because the earlier predicted $\alpha_q$ values are obtained using the non-relativistic methods or lower-order relativistic methods. Thus, only measurements can only ascertain reliability of these calculations. Now comparing the $\alpha_d$ value of Ne with experiment \cite{expt1}, we find that our RCCSD value is more close with the experiment than the RNCCSD value. We anticipate that after including Breit and quantum electrodynamics corrections, the RNCCSD value will improve further. Similarly, the $\alpha_q$ value from the RCCSD method is found to be more close with the previous calculations than the RNCCSD method. Since there is no experimental result for $\alpha_q$ available, we cannot claim that the RNCCSD value is less accurate than RCCSD result. Similar trends for the $\alpha_d$ and $\alpha_q$ values can be seen in the Ar atom.

Compared to the $\alpha_d$ and $\alpha_q$ values, $B$ values have got a little attention both in the theoretical and experimental studies. Contributions from these values are extremely small to the Stark effects to being observed precisely. Strong electron correlation effects are also involved in evaluating $B$ values accurately. In addition, extrapolation of $B$ values from the FF approach requires inclusion of both the electric dipole and quadrupole field interactions in the atomic Hamiltonian. In the linear response approach, estimations of $B$ values demand calculation of first-order perturbed wave functions due to both the electric dipole and quadrupole operators. These are the main reasons why $B$ values are not widely investigated in many atomic systems. We found a few literature values for $B$ of the Be, Mg, Ne and Ar atoms \cite{thakkar,ivan,thar,taylor}, which are listed in Table \ref{tab1}. These literature values are basically obtained by adopting FF approach in the non-relativistic framework. From the comparison with our calculations with the literature values, we find that our RNCCSD values agree with the earlier reported values while the RCCSD results differ a lot in both the Be and Mg atoms. However, they are other way around for the Ne and Ar atoms. The reasons for this could be different many-body methods were considered to estimate $B$ values of the alkaline-earth atoms and of the noble gas atoms. We expect that our RNCSSD results are more reliable compared to all the listed values of Table \ref{tab1}.

In Figs. \ref{potbemg}(a) and (b), we show individual contributions to $V(r)$ from the RCCSD and RNCCSD methods for the Be atom. As can be seen in both these plots, contributions from $V_{nuc}$ dominants while $V_{C}$ contributions are also quite visible. There are also noticeable contributions arising from the $V_{pol}(r)$. Similar trends can also be observed from the RCCSD values, shown in Fig. \ref{potbemg}(c), and the RNCCSD values, shown in Fig. \ref{potbemg}(d), for the Mg atom but the shapes are slightly different due to $V_C(r)$ and $V_{pol}(r)$ contributions. These differences can be understood from the density profiles of both the atoms shown in Fig. \ref{densfig}. We show different contributions to $V(r)$ from the RCCSD and RNCCSD methods for Ne and Ar in Figs. \ref{potnear}(a)-(d). As can be seen from the figure, the trends from individual contribution to $V(r)$ in both Ne and Ar almost look similar except in their magnitudes. It also shows that contributions from $V_{pol}(r)$ are negligibly small in both the atoms. Compared to the alkaline-earth atoms, results for both the Ne and Ar atoms look slight similarity with the Mg atom. It is worth noting that the density profiles between the Be and Ar atoms looked to be similar while density profiles between the Mg and Ne seemed to have similar features in Fig.  \ref{densfig}. Thus, it is not possible to get clear picture of the scattering potential behavior of an electron from an atom by just looking at the density profile of the atom. Nonetheless, we have discussed procedures to construct the  electron-atom scattering potentials by evaluating contributions from the static and polarization potentials due to the Be, Mg, Ne and Ar atoms using the RCC and RNCC methods. These procedures can also be adopted in the heavier closed-shell atomic systems, where the electron correlation effects could be very much pronounced.
 
\section{Conclusions}\label{conclusions}

We have demonstrated approaches to employ relativistic-coupled cluster theory to determine potentials for the evaluation of electron-atom elastic scattering cross-sections. For this purpose, we have considered both the standard and normal version of the relativistic coupled-cluster theory in the singles and doubles approximation, and presented results for the Be, Mg, Ne and Ar atoms as representative elements for the alkaline-earth and noble gas atoms of the periodic table. To estimate the static potential contributions, finite-size nuclear effect has been accounted through the nuclear potential while the two-electron correlation effects are estimated using the relativistic coupled-cluster theory. Density functions of the above atoms from both the considered relativistic coupled-cluster theories are presented in order to estimate the Coulomb exchange potential contributions, which we have neglected in this work for estimating potentials. Furthermore, we have determined the electric dipole, quadrupole and dipole-quadrupole polarizabilities to account for the electron polarization effects to the scattering potential. Results from both the standard and normal relativistic coupled-cluster theories are compared with the literature values. These methods can be further applied to other heavier atomic systems to study electron-atom scattering cross-sections more accurately where electron correlation effects within the atom will be more prominent than the presently investigated lighter elements.

\section*{Acknowledgement}
 
We acknowledge use of Vikram-100 HPC cluster of Physical Research Laboratory (PRL), Ahmedabad, India for the computations.

\end{document}